\documentclass[pdflatex,sn-mathphys]{sn-jnl}% Math and Physical Sciences Reference Style
\usepackage[utf8]{inputenc}
\usepackage{slashed}
\hypersetup{colorlinks=false}
\usepackage{lineno}
\usepackage{color}
\definecolor{oldtxtcolor}{rgb}{0.00, 0.0, 0.5}
\definecolor{newtxtcolor}{rgb}{0.0, 0.0, 0.0}
\definecolor{newtxtcolor2}{rgb}{0.5, 0.0, 0.5}
\definecolor{newtxtcolor3}{rgb}{0.00, 0.0, 0}
\definecolor{newtxtcolor4}{rgb}{0.0, 0.0, 0.0}
\definecolor{newtxtcolor5}{rgb}{0.0, 0.0, 0.0}
\definecolor{newtxtcolor6}{rgb}{0.0, 0.0, 0.0}
\definecolor{oldtxtcolor}{rgb}{1.00, 0.0, 0.00}

\def\verX{12}
\def\verO{1}
\def\verN{2}
\def\verON{12}

\ifx\verX\verO
 \newcommand { \oldtxt }[1] {{\color{oldtxtcolor}{#1}}}
 \newcommand { \newtxt }[1] {}
\fi
\ifx\verX\verN
 \newcommand { \oldtxt }[1] {}
 \newcommand { \newtxt }[1] {{\color{newtxtcolor}{#1}}}
\fi
\ifx\verX\verON
 \newcommand { \oldtxt }[1] {{\color{oldtxtcolor}{#1}}}
 \newcommand { \newtxt }[1] {{\color{newtxtcolor}{#1}}}

\fi

\jyear{2022}

\begin{document}
%\linenumbers
\title{Entangled X-ray Photon Pair Generation by Free Electron Lasers}


\author[1]{\fnm{Linfeng} \sur{Zhang}}%\email{@}
\equalcont{These authors contributed equally to this work.}

\author[1]{\fnm{Zunqi} \sur{Li}}%\email{@}
\equalcont{These authors contributed equally to this work.}

\author[1,6]{\fnm{Dongyu} \sur{Liu}}%\email{@}

% \author[1]{\fnm{Ming} \sur{Zhang}}%\email{@}

\author[1,4,5]{\fnm{Chengyin} \sur{Wu}}%\email{@}

%\author[1]{\fnm{Yunquan} \sur{Liu}}%\email{@}

%\author*[1,3,4]{\fnm{Qihuang} \sur{Gong}}\email{qhgong@pku.edu.cn}

\author*[2,3]{\fnm{Haitan} \sur{Xu}}\email{haitanxu@nju.edu.cn}

\author*[1,4,5]{\fnm{Zheng} \sur{Li}}\email{zheng.li@pku.edu.cn}

\affil[1]{\orgdiv{State Key Laboratory for Mesoscopic Physics and Collaborative Innovation Center of Quantum Matter, School of Physics}, \orgname{Peking University}, \orgaddress{\city{Beijing}, \postcode{100871}, \country{China}}}

\affil[2]{\orgdiv{School of Materials Science and Intelligent Engineering}, \orgname{Nanjing University}, \orgaddress{\city{Suzhou}, \postcode{215163}, \state{Jiangsu}, \country{China}}}
\affil[3]{\orgdiv{Shenzhen Institute for Quantum Science and Engineering}, \orgname{Southern University of Science and Technology}, \orgaddress{\city{Shenzhen}, \postcode{518055}, \state{Guangdong}, \country{China}}}

\affil[4]{\orgname{Peking University Yangtze Delta Institute of Optoelectronics}, \orgaddress{\city{Nantong}, \country{China}}}

\affil[5]{\orgdiv{Collaborative Innovation Center of Extreme Optics}, \orgname{Shanxi University}, \orgaddress{\street{}, \city{Taiyuan}, \postcode{030006}, \state{Shanxi}, \country{China}}}

\affil[6]{\orgdiv{Physics Department}, \orgname{Stanford University}, \orgaddress{\city{CA}, \postcode{94305}, \country{USA}}}

\maketitle
\pagestyle{plain}

%Abstract, 142/180 words
\textbf{
Einstein, Podolsky and Rosen's prediction~\cite{einstein1935can} on incompleteness of quantum mechanics was overturned by experimental tests on Bell's inequality~\cite{bell1966problem} that confirmed the existence of quantum entanglement. In X-ray optics, entangled photon pairs can be generated by X-ray parametric down conversion (XPDC), which has certain wavelength window~\cite{shwartz2012x}. Meanwhile, free electron laser (FEL) has successfully lased at X-ray frequencies recently~\cite{emma2010first,ishikawa2012compact,allaria2013two,wang2021free,pompili2022free,duris2020tunable,duris2021controllable}. However, FEL is usually seen as a classical light source, and its quantum effects are considered minor corrections to the classical theory. 
Here we investigate entangled X-ray photon pair emissions in FEL. We establish a theory for coherently amplified entangled photon pair emission from microbunched electron pulses in the undulator. \newtxt{We numerically demonstrate the properties of entangled emission, and provide a scheme to generate highly entangled X-ray photon pairs, which is of great importance in X-ray quantum optics. Our work shows a unique advantage of FELs over synchrotrons in entangled X-ray photon pair generation.}
} 

Being an ultrabright and ultrashort X-ray laser source, FEL has shed light on various research areas including single particle imaging~\cite{seibert2011single,loh2012fractal,gomez2014shapes}, ultrafast X-ray spectroscopy~\cite{kern2013simultaneous,erk2014imaging,rudek2012ultra} and high energy density physics~\cite{vinko2012creation,fletcher2015ultrabright,ciricosta2016measurements}. The FEL emission was mostly seen as a classical phenomenon with minor quantum corrections~\cite{pellegrini2016physics}.
However, a critical quantum effect, i.e., the quantum entanglement, has never been investigated for the X-ray emission from the FEL. 

In this work, we study the entangled X-ray photon pair emission from FEL, which can be enhanced by electron microbunching in the undulator. We start by presenting the quantum electrodynamics (QED) treatment of single-electron photon pair emission process in the undulator and calculate the cross section and the entanglement degree. Then we apply the Feynman rules to microbunched electrons, which takes into account the many-body effect, to analyze the coherent amplification and establish the condition for coherence. \newtxt{We further numerically investigate the enhancement of entangled photon pair emission by microbunched electrons.}

\section*{Entangled photons from FEL}

\begin{figure*}[thb]
    \centering
    \includegraphics[width = \textwidth]{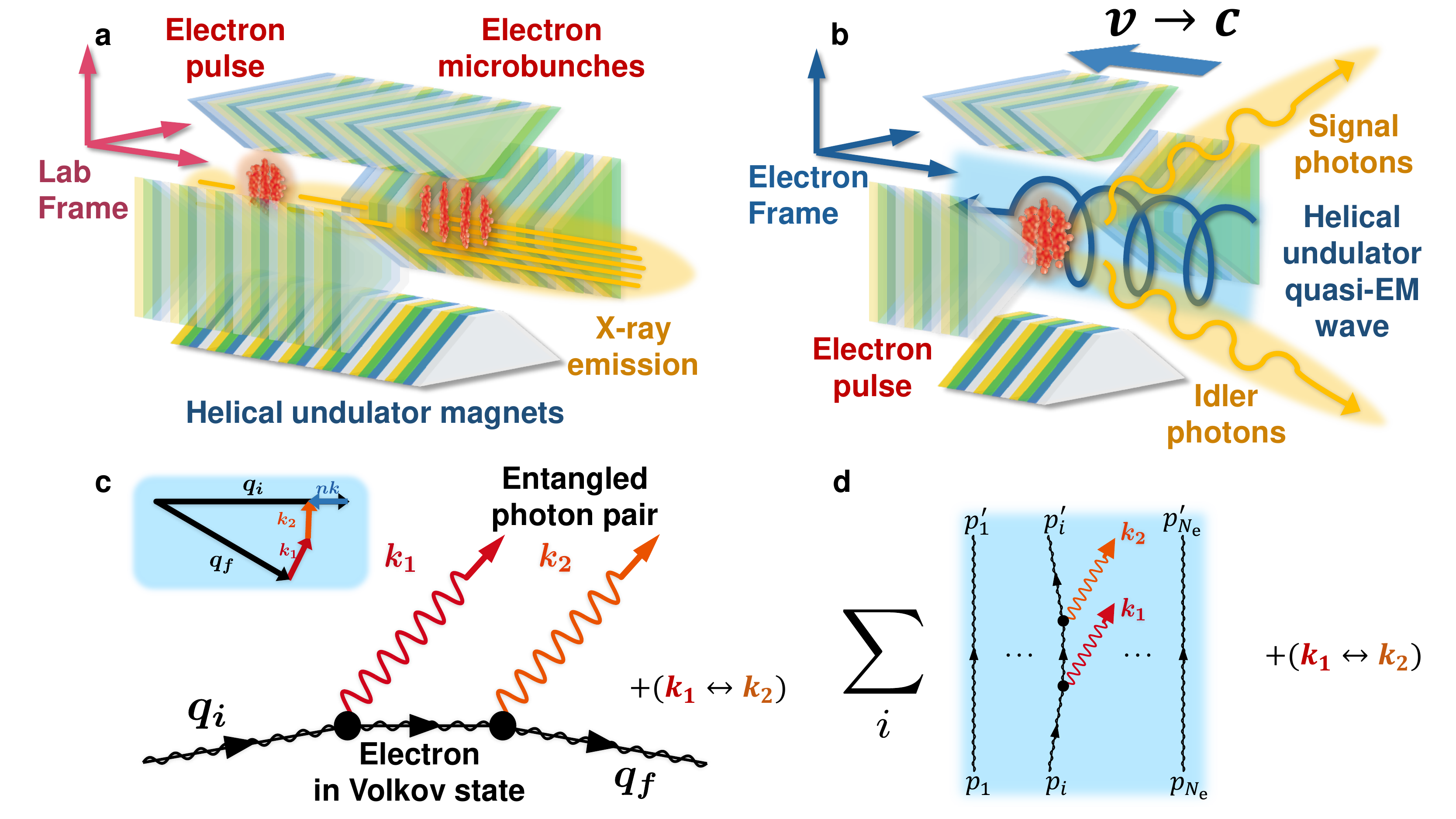}
    \caption{\textbf{Mechanism of coherent entangled X-ray photon pair emission in SASE FEL}. \textbf{a}, Schematic in the lab frame, where a relativistic electron pulse (red) passes through an undulator, causing photon emission (yellow) while being microbunched. The fundamental frequency of the emission can reach X-ray region, with single-photon energy of hundreds of $\mathrm{eV}$s. The schematic of the helical undulator is drawn according to the configuration of the Delta undulator at LCLS~\cite{lutman2016polarization}. \textbf{b}, Schematic in the electron frame, where the electron pulse is initially at rest. The strong quasi-EM wave (blue) of the relativistic undulator  is scattered from the electrons. The resulting emission contains multiple components, including single-photon emission, double-photon  emission, etc. \textbf{c}, The Feynman diagrams corresponding to double-photon emission in the electron frame. The black lines refer to the Volkov states of electron. Inset: a typical constraint of four-momenta of the particles. $n$ is the net number of Floquet photons absorbed. \textbf{d}, Coherent summation of Feynman diagrams with $n=1$, including permutations of $k_1$ and $k_2$.}
    \label{fig:Pano}
\end{figure*}

We present in Fig.~\ref{fig:Pano} the quantum mechanism of entangled X-ray photon pair emission. Fig.~\ref{fig:Pano}a shows a typical SASE FEL process in the lab frame. A relativistic electron pulse travels through an undulator, interacts with the electromagnetic (EM) field of the undulator, resulting in electron microbunching and amplified emission. 

To facilitate the study of entangled photon emission from the FEL, we use the electron frame (EF), where the incident electron pulse has a zero initial average velocity. We adopt the frame transformation formalism from the Weizs{\"a}cker-Williams  method~\cite{madey1971stimulated}, and use the natural units $\hbar=c=\epsilon_0=1$. For a helical undulator, $A_{\mathrm{Lab}}^{\mu}=a(0,\mathrm{cos}(k_ux_{\mathrm{Lab}}^3),\mathrm{sin}(k_ux_{\mathrm{Lab}}^3),0)$, where $a=B_0/k_u$, $B_0$ is the amplitude of the undulator field, and $k_u=2\pi/\lambda_u$ is the spatial frequency of the undulator with a period of $\lambda_u$ in the lab frame~\cite{temnykh2008delta}. The undulator field is transformed to a quasi-EM wave in the EF by the Lorentz boost as

\begin{equation}
    \begin{split}\label{eq:A}
        A_{\mathrm{EF}}^{\mu}&=a(0,\mathrm{cos}[k_u(\gamma\beta x_{\mathrm{EF}}^0+\gamma x_{\mathrm{EF}}^3)], \\&\phantom{=}\mathrm{sin}[k_u(\gamma\beta x_{\mathrm{EF}}^0+\gamma x_{\mathrm{EF}}^3)],0) \\
        &=a(0,\mathrm{cos}(k\cdot x_{\mathrm{EF}}),\mathrm{sin}(k\cdot x_{\mathrm{EF}}),0)\;,
    \end{split}
\end{equation}
where $\beta$ and $\gamma$ are the average velocity and Lorentz factor of the electron pulse. $k=\gamma k_u(\beta,0,0,-1)$ is the wave vector of the undulator quasi-EM wave in the EF, which is close to the photon mass shell for $\beta\to 1$. The helical undulator is practically available, such as the Delta undulator at LCLS~\cite{lutman2016polarization}.

The SASE emission can be approximated as the scattering of the quasi-EM wave from the electron pulse in the EF. The emission takes part in the microbunching of the electron pulse, and is significantly enhanced at the fundamental frequency and its harmonics. In the lab frame, the fundamental frequency $\omega_{\mathrm{fd}}=2\gamma^2(2\pi/\lambda_u)/(1+K^2)$ for the helical undulator, where $K=eB_0/mk_u=ea/m$ is the undulator parameter~\cite{emma2010first, pellegrini2016physics, huang2007review}.

For strong undulator quasi-EM wave, nonlinear scattering can be observed in the emission. We focus on the double-scattering processes, as shown in Fig.~\ref{fig:Pano}b. In this work we use the FEL parameters of Linac Coherent Light Source (LCLS)~\cite{emma2010first}, as shown in Table~\ref{tab:ExpPara}.

For a representative electron initially at rest in the EF, it satisfies the Dirac equation $(\slashed{p}-e\slashed{A}-m)\Psi=0$, where  $\slashed{f}=\gamma^{\mu}f_{\mu}$, $\gamma^{\mu}$ are the Dirac matrices, $A$ is the EM four-potential, $e$ is the electron charge, and $m$ is the rest electron mass. 

% \begin{table*}[htb]
%     \centering
%     \begin{tabular}{cccc}
%         \hline\hline
%         Electron energy & $5~\mathrm{GeV}$ & Pierce parameter & $2\times 10^{-3}$ \\
%         Charge per pulse & $180~\mathrm{pC}$ & Fundamental wavelength & $2.30~\mathrm{nm}$ \\
%         Peak current &  $3.4~\mathrm{kA}$ & Saturation peak power & $10~\mathrm{GW}$ \\
%         Repetition rate & $120~\mathrm{Hz}$ & Emission pulse FWHM & $230~\mathrm{fs}$ \\
%         Undulator peak field & $1.32~\mathrm{T}$ & Non-linearity parameter $\xi$ & $3.70$ \\
%         Undulator period length & $30~\mathrm{mm}$ & Quantum parameter $\chi$ & $2.9 \times 10^{-6}$ \\
%         Undulator period number & $1392$ & & \\
%         \hline
%     \end{tabular}
%     \caption{FEL parameters used in this work, based on LCLS~\cite{emma2010first}. The non-linearity parameter $\xi=ea/m$ and the quantum parameter $\chi=\gamma\xi\lambda_{\mathrm{C}}/\lambda_{\mathrm{u}}$ reflect the relative magnitudes of the nonlinear effect to the linear effect and the nonlinear quantum effect to the nonlinear classical effect, respectively. $\lambda_{\mathrm{C}}$ is the Compton wavelength~\cite{ritus1985quantum,di2012extremely}. }
%     \label{tab:ExpPara}
% \end{table*}

\begin{table*}[htb]
    \centering
    \begin{tabular}{cccc}
        \hline\hline
        Electron energy & $5~\mathrm{GeV}$ & Undulator period number & $1392$ \\
        Charge per pulse & $180~\mathrm{pC}$ & Pierce parameter & $2\times 10^{-3}$ \\
        Peak current & $3.4~\mathrm{kA}$ & Fundamental wavelength & $2.30~\mathrm{nm}$ \\
        Repetition rate & $120~\mathrm{Hz}$ &  Saturation peak power & $10~\mathrm{GW}$ \\
        Undulator peak field & $1.32~\mathrm{T}$ & Emission pulse FWHM & $230~\mathrm{fs}$ \\
        Undulator period length & $30~\mathrm{mm}$ & & \\
        \hline
    \end{tabular}
    \caption{FEL parameters used in this work, based on LCLS~\cite{emma2010first}.}
    \label{tab:ExpPara}
\end{table*}

In the undulator quasi-EM wave, the electron is in a Volkov state $\Psi^{\mathrm{V}}_p(x)=(1-\frac{e\slashed{A}\slashed{k}}{2k\cdot p})u_{p}\times\mathrm{exp}\{-i[p\cdot x+\int_{0}^{\phi}{(\frac{ep\cdot A(\phi')}{k\cdot p}-\frac{e^2 A^2(\phi')}{2k\cdot p})d\phi'}]\}$, where  
$\phi=k\cdot x=k^{\mu}x_{\mu}=k^0x^0-\boldsymbol{k}\cdot\boldsymbol{x}$,  $u_{p}$ is the free electron spinor, and $A\simeq A_{\mathrm{EF}}$ [see Section I of Supplementary Information (SI) for details]. For an electron of initial four-momentum $p$, the corresponding Volkov state has a time-varying four-momentum whose average is the quasi-momentum $q^{\mu}=p^{\mu}+\frac{e^2a^2}{2k\cdot p}k^{\mu}$, satisfying $\vert q\vert^2=m^2+e^2a^2=m_*^2$. 

In the Floquet picture, the Volkov state corresponds to the superposition state of the electron dressed by $\tilde{n}$ Floquet photons, i.e., $\Psi^{\mathrm{V}}_p(x)=\sum_{\tilde{n}=-\infty}^{+\infty}{F_{\tilde{n}} \mathrm{exp}\{-i(q\cdot x+\tilde{n} k\cdot x)}\}u(p)$ (the expression of $F_{\tilde{n}}$ can be found in SI). The electron in the Volkov state can spontaneously decay into another Volkov state with a different quasi-momentum and thus radiate photons, the amplitude of which can be evaluated by the Feynman diagrams (see details in Section II of SI). Especially, the entangled photon pair emission is equivalent to non-perturbative double Compton scattering of the quasi-EM wave from the electron in the EF~\cite{ritus1985quantum,lotstedt2009nonperturbative}, which satisfies the four-momentum conservation $q_\mathrm{f}+k_1+k_2-q_\mathrm{i}-nk=0$ (see Fig.~\ref{fig:Pano}c). $k_1$ and $k_2$ are the momentum of entangled photons, and $n$ corresponds to the number of Floquet photons absorbed. As the energies of photons ($k^0,k_1^0,k_2^0$) are much less than the electron mass ($m$) and the quasi-momenta ($q_{\mathrm{i}},q_{\mathrm{f}}$) in the EF when $n$ is reasonably small, the four-momentum conservation leads to $k_1^0+k_2^0\simeq nk^0$. We focus on the representative case of $n=1$, which  corresponds to photon pairs with total frequency of  $\omega_{\mathrm{fd}}$ in the lab frame, and can be distinguished from other cases with different $n$ by detecting the emitted photon pair energies (see details in Section III of SI). 

With the above treatment, we obtain the scattering matrix element $S_{\mathrm{fi}}$ for the process from an initial Volkov state $\mathrm{i}$ to any final Volkov state $\mathrm{f}$  emitting a pair of photons in plane wave states with arbitrary polarization or helicity. \newtxt{The differential double-photon emission rate is given by
\begin{equation}\label{eq:ddotW}
    \begin{split}
        &\frac{d\dot{W}}{dk_1^0d\Omega_{k_1}d\Omega_{k_2}}=\frac{e^4}{(2\pi)^5}\frac{m^2}{2q_{\mathrm{i}}^0} \\
        &\phantom{=}\times\sum_{n=1}^{+\infty}{\frac{k_1^0(k_2^0)^2\Theta(q_{\mathrm{i}}^0+nk^0-k_1^0-k_2^0)M}{2\vert(q_{\mathrm{i}}+nk-k_1)\cdot k_2\vert}}\bigg\vert_{\mathrm{DE}(n)}\;,
    \end{split}
\end{equation}
where $\Omega_{k_1}$ and $\Omega_{k_2}$ are the solid angles of the emitted photon pair, the notation $\vert_{\mathrm{DE}(n)}$ represents the constraint of $ (q_{\mathrm{i}}+nk-k_1)\cdot k_2=q_{\mathrm{i}}\cdot nk-q_{\mathrm{i}}\cdot k_1-nk\cdot k_1$, and the helicity-relevant part $M$ is explained in SI (see details in Methods and Section II of SI). }

\newtxt{In order to quantify the entanglement degree of the emitted X-ray photon pairs, we calculate the density matrix $\rho_{\mathrm{f}}$ in the helicity basis from the scattering matrix elements $S_{\mathrm{fi}}$ as} $\rho_{\mathrm{f}}=\frac{1}{2}\sum_{r_{\mathrm{i}},r_{\mathrm{f}}=1}^{2}N\big(S_{\mathrm{fi},j_1}S_{\mathrm{fi},j_2}^*\big)_{4\times4}$, where $S_{\mathrm{fi},j_1}$ and $S_{\mathrm{fi},j_2}$ represent the scattering matrix elements with the emitted photon pair at helicity eigenstates $j_1$ and $j_2$, and $j_1,j_2=1,2,3,4$ correspond to the helicity eigenstates $\vert ++ \rangle,\vert +- \rangle,\vert -+ \rangle$, and $\vert -- \rangle$, where $\pm$ corresponds to the right/left-handed helicity~\cite{lotstedt2009correlated}. $N$ is chosen to keep $\mathrm{Tr}[\rho_{\mathrm{f}}]=1$. $r_{\mathrm{i}}$ and $r_{\mathrm{f}}$ represent the spin of the electron at initial and final states, respectively, which have been traced out. The scattering matrix elements can be calculated for any choice of $k_1,k_2$ satisfying the constraint $(k_1+k_2-q_{\mathrm{i}}-nk)^2-m_*^2=0$. 

In order to characterize the quantum entanglement of the emitted photon pair, we use the so-called concurrence as a measure of the entanglement~\cite{wootters1998entanglement}. The concurrence can be calculated from the density matrix as

\begin{equation}
    \mathcal{C}(\rho_{\mathrm{f}})=\mathrm{max}(0,\sqrt{\zeta_1}-\sqrt{\zeta_2}-\sqrt{\zeta_3}-\sqrt{\zeta_4})\;,
\end{equation}
where $\zeta_{1,2,3,4}$ are the 4 eigenvalues of the matrix $Q=\rho_{\mathrm{f}}(\sigma^2\otimes\sigma^2)\rho_{\mathrm{f}}^*(\sigma^2\otimes\sigma^2)$ in descending order, and $\sigma^2$ is one of the Pauli matrices. $\mathcal{C}\in[0,1]$, with $\mathcal{C}=0$ corresponding to a unentangled state and $\mathcal{C}=1$ corresponding to a fully entangled state.
As the helicity of massless particles like photons are Lorentz invariant, the density matrix and the concurrence are also Lorentz invariant.  

\newtxt{We have represented the density matrix and the concurrence of the 2-qubit states of the emitted photon pairs with the basis of the helicity eigenstates $\vert ++ \rangle,\vert +- \rangle,\vert -+ \rangle,\vert -- \rangle$. The choice of basis is not unique, and we can alternatively choose other basis states, such as linear polarization eigenstates~\cite{lotstedt2009correlated}, which result in a unitary transformation of the density matrix, and the concurrence remains invariant (see Section VII of SI for details). In addition, we can choose other entanglement monotones instead of the concurrence to quantify the entanglement degree, such as entanglement of formation and negativity. A comparison showing qualitative agreement between these entanglement measures and the concurrence is given in Section IX of SI.}

\section*{Microbunch enhancement}

In the high-gain regime of FEL, a coherent enhancement due to the electron microbunching plays a vital role. For a sufficiently bright electron beam and a sufficiently long undulator, the combination of the undulator field and the radiation field will induce a ponderomotive potential modulating the energy of the electron beam~\cite{huang2007review}. This energy modulation eventually forces the electrons to form periodic microbunches along the undulator axis with a modulation wavelength $\lambda_{1}=2\pi /\omega_{\mathrm{fd}}$ equal to the fundamental wavelength. 

The conventional analysis of the coherence property due to the microbunching relies on paraxial wave equation~\cite{huang2007review}, which cannot explain the quantum phenomenon of entangled photon pair emission. To overcome this problem, we develop a quantum collective double emission (QCDE) theory (see details in Section IV of SI). \newtxt{This theory provides a QED description of the microbunched state of electrons, and deals with the interference of the Feynman diagrams corresponding to double emissions from different electrons. The collective double emission rate in the QCDE theory is
\begin{equation}\label{eq:dWMb}
    \begin{aligned}
        \bigg(\frac{d\dot{W}}{dk_1^0d\Omega_{k_1}d\Omega_{k_2}}\bigg)_{\mathrm{C}}=\frac{d\dot{W}}{dk_1^0d\Omega_{k_1}d\Omega_{k_2}}F_{\mathrm{MB}}(Z_1',Z_2',k_1^0)\;,
    \end{aligned}
\end{equation}
where $F_{\mathrm{MB}}(Z_1',Z_2',k_1^0)$ is the enhancement factor on the scattering rate introduced by microbunching. $(d\dot{W}/dk_1^0d\Omega_{k_1}d\Omega_{k_2})_{\mathrm{C}}$ represents the collective differential scattering rate for a microbunched electron pulse.}
\newtxt{In addition to the rigorous QCDE theory presented in SI, here we elaborate the physical picture leading to the coherent enhancement factor $F_{\mathrm{MB}}(Z_1',Z_2',k_1^0)$ via an analogy with the quasi-phase-matching in nonlinear optics (see details in Section IV of SI). }

In the electron frame, we can view the quasi-EM wave generated by the relativistic undulator as the pump and the emitted photon pair as  signal and  idler. The electron microbunches can be regarded as a nonlinear medium with periodic structure. Therefore, the phase difference between photons emitted by adjacent microbunches should be equal to $\Delta\phi=\Delta\boldsymbol{l}\cdot \Delta\boldsymbol{k}$, where $\Delta\boldsymbol{l}=\vert\Delta\boldsymbol{l}\vert \hat{\boldsymbol{e}}^3=(\lambda_u/\gamma)(1+K^2)/(2+K^2)\vert \hat{\boldsymbol{e}}^3$ is the average relative displacement between adjacent microbunches, with $\hat{\boldsymbol{e}}^3$ being the unit vector along the undulator, and $\Delta\boldsymbol{k}=\boldsymbol{k_1}+\boldsymbol{k_2}-n\boldsymbol{k}$ is the momentum transfer between the pump photon and the emitted photon pair.  

If the momenta of the signal and idler photons are both along the $x^3$ axis, the total momentum of the photon pair is $\boldsymbol{k_1}+\boldsymbol{k_2}=-n\boldsymbol{k}/(1+K^2)$. Therefore the phase difference is exactly $\Delta\phi=2\pi n$, resulting in a fully constructive interference between emissions from different electron microbunches and hence an $N_{\mathrm{e}}^2$ enhancement of the collective emission rate.
In a practical FEL system, the situation deviates from the ideal one, and only adjacent  microbunches within the coherence length contribute to the constructive interference, the number of which  is estimated to be about $N_\mathrm{c}=22$ for the parameters shown in Table~\ref{tab:ExpPara}. 

Meanwhile, if the directions of the emitted photon pair are away from the $x^3$ axis, the phase difference will deviate from $2\pi n$ with a nontrivial angular dependence

\begin{equation}
    \begin{split}
        \Delta\phi(Z_1',Z_2',k_1^0)&=2\pi\Big[(1+K^2) \\
        &\phantom{=}\times\frac{1+\mathrm{cos}(Z_2')-(\mathrm{cos}(Z_2')-\mathrm{cos}(Z_1'))k_1^0/k^0}{2+K^2(1+\mathrm{cos}(Z_2'))}\Big]
    \end{split}
\end{equation}
(see details in Section IV of SI), where $Z_1',Z_2'$ are the zenith angles  of the two emitted photons relative to the $x^3$ axis in  the electron frame. \newtxt{The enhancement parameter on the scattering amplitude by electron microbunches within one coherent length is given by $H(Z_1',Z_2',k_1^0)=\sum_{j=1}^{N_{\mathrm{c}}}N_j e^{i[j\Delta\phi(Z_1',Z_2',k_1^0)]}$, where $N_{j}$ is the number of electrons within the ${j}^{\mathrm{th}}$ electron microbunch. The collective emission rate of such coherent microbunches is the product of the emission rate from a single electron and the enhancement parameter $\vert H(Z_1',Z_2',k_1^0)\vert ^2$. For the entire electron pulse,  the coherent enhancement factor in Eq.~\ref{eq:dWMb} is given by
\begin{equation}\label{eq:dWM}
    \begin{aligned}
        F_{\mathrm{MB}}=\vert H(Z_1',Z_2',k_1^0)\vert ^2 N_\mathrm{i}\;,
    \end{aligned}
\end{equation}
where $N_\mathrm{i}\simeq N_{\mathrm{e}}/(\sum_{j=1}^{N_{\mathrm{c}}}N_j)$ represents the approximate number of coherent sections.
}

\newtxt{In addition, using the Lorentz transformation $\gamma\mathrm{tan}(Z_l)=\mathrm{sin}(Z_l')/(\beta+\mathrm{cos}(Z_l'))$, we can determine the zenith angles $Z_1$ and $Z_2$ in the lab frame for observation of the entangled photon pair emission. 
%
%For the entanglement degree of the emitted photon pairs from microbunched electrons, 
As the density matrix of the emitted photon pairs from microbunched electrons has an identical form as that from the single electron emission, the entanglement degree is retained under microbunching enhancement.}
%i.e., every element of the density matrix can be seen as the corresponding element in single electron density matrix multiplying the same factor $F_{\mathrm{MB}}$, which is always cancelled in normalization of the density matrix.

\section*{Numerical results}

\begin{figure*}[thb]
    \centering
    \includegraphics[width = \textwidth]{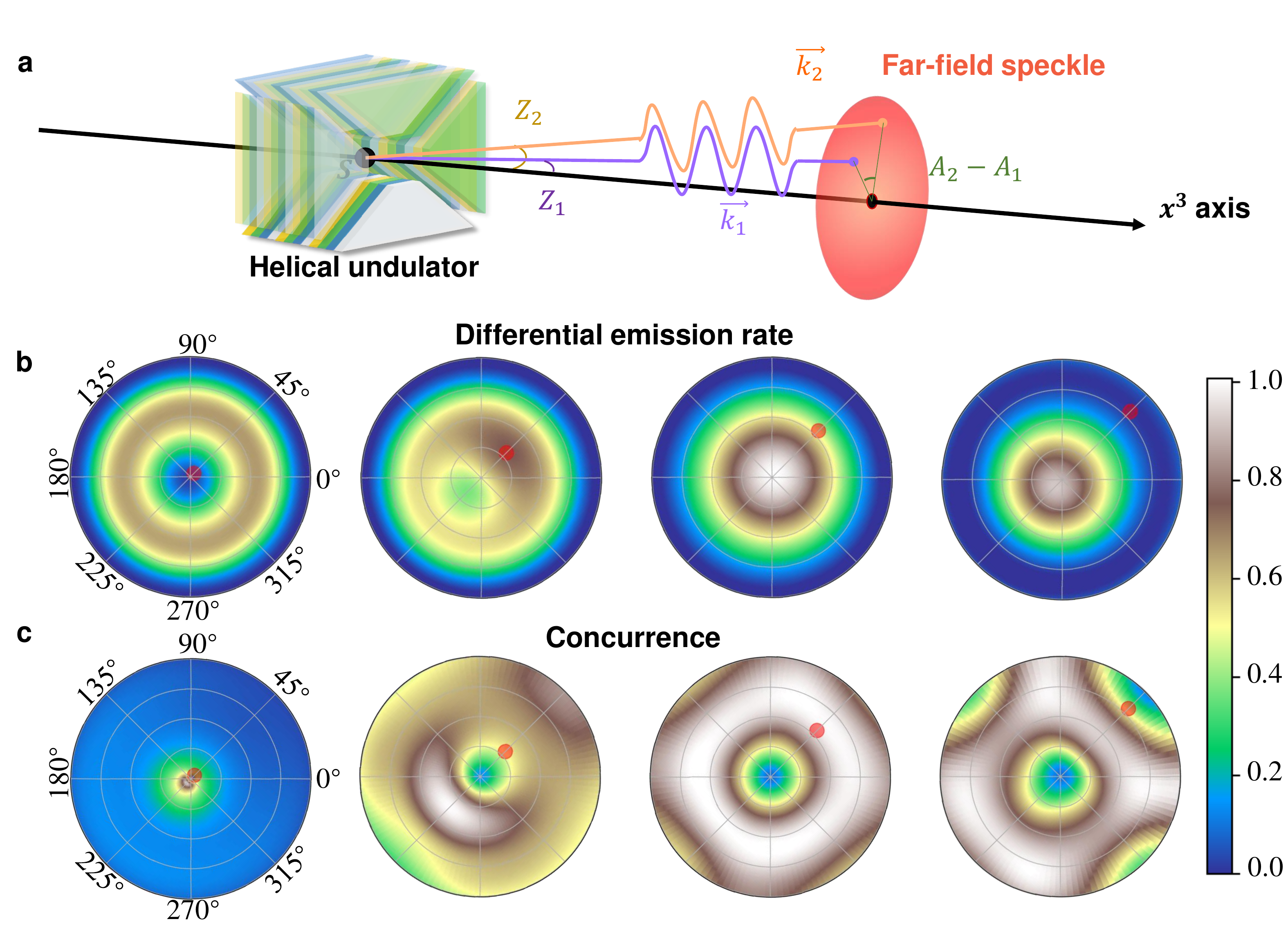}
    \caption{\textbf{Numerical simulation of entangled X-ray photon pair emission with microbunch enhancement.}  \textbf{a}, Illustration of the angles in \textbf{b} and \textbf{c}. $Z_{1,2}$ refer to the zenith angle between the three-momentum $\boldsymbol{k}_1,\boldsymbol{k}_2$ of the emitted photons and the $x^3$ axis. $A_2-A_1$ is the difference between the azimuth angles of the emitted photons at the transverse far-field plane. Photons are treated as being emitted from the point $S$ at the center of undulator for far-field measurement. \textbf{b}, Normalized double-angular distribution of the differential scattering rate $d\dot{W}/dk_1^0d\Omega_{k_1}d\Omega_{k_2}$ in the lab frame. \textbf{c}, Double-angular distribution of the concurrence $\mathcal{C}$. 
    In both \textbf{b} and
    \textbf{c}, the polar angles refers to $A_2\in[0,2\pi]$, and the radii refers to $\gamma\mathrm{tan}(Z_2)\in[0,1]$ for photon 2. The red points represent the directions of photon 1.
    % In \textbf{b} and \textbf{c}, we display the double-angular distribution at a far-field speckle in polar figures for specified directions of photon 1 bearing $k_1$, whose position $\gamma\mathrm{tan}(Z_1),~A_1=0$ is labelled with the red circle. Meanwhile, the distribution in each polar figure is on the direction of photon 2 bearing $k_2$, with the radii referring to $\gamma\mathrm{tan}(Z_2)\in[0,1]$, and the polar angles referring to $A_2-A_1\in[0,2\pi]$.
    Here we choose $\omega_1\simeq\omega_{\mathrm{fd}}/3$.}
    \label{fig:Num_Res}
\end{figure*}

%\newtxt{Here we make numerical calculations based on the theory above}. 
We firstly calculate the total scattering rate $\dot{W}_{\mathrm{fi}}$. We set practical ranges of photon energies and angular directions for both photon detectors, and focus on the case of $n=1$. The relevant angles are illustrated in Fig.~\ref{fig:Num_Res}a. For the helical undulator, the dependence on the azimuthal angles $A_1,A_2$ of the emitted photon pair is simplified to  $A_2-A_1$. Thus, $d\dot{W}/dk_1^0d\Omega_{k_1}d\Omega_{k_2}$ can be displayed in a three-dimensional space of the zenith angles $Z_1,Z_2$ and $A_2-A_1$. Choosing $\omega_1\simeq\omega_{\mathrm{fd}}/3$ and $\omega_2\simeq2\omega_{\mathrm{fd}}/3$ in the lab frame, the double angular distribution of the differential scattering rate $d\dot{W}/dk_1^0d\Omega_{k_1}d\Omega_{k_2}$ is shown in Fig.~\ref{fig:Num_Res}b.
%After a weighed sum over solid angle $\Omega_{k_1},\Omega_{k_2}$ and an energy range of $30~\mathrm{eV}$, the angular integrated total scattering rate by all electrons is about $4\times10^{14}~\mathrm{s}^{-1}$ to $2\times10^{-1}~\mathrm{s}^{-1}$. The traveling time through the undulator in the lab frame is $139~\mathrm{ns}$ for the parameters in Table~\ref{tab:ExpPara}. Thus, for one electron pulse passing the undulator, we can create $6\times10^{7}$ to $3\times10^{-4}$ photon pairs. 
%After a weighed sum over the solid angles $\Omega_{k_1},\Omega_{k_2}$, with an energy range of $30~\mathrm{eV}$  and the traveling time through the undulator shown in Table~\ref{tab:ExpPara}. 
%, the angle-integrated total scattering rate for one electron pulse is $4\times10^{-2}~\mathrm{s}^{-1}$. }

To demonstrate the quantum entanglement, we show the concurrence $\mathcal{C}$ of the emitted photon pairs in Fig.~\ref{fig:Num_Res}c. Especially, $\mathcal{C}$ can reach its upper limit of $1$ for certain angles, meaning that fully-entangled two-photon states may be generated. Concurrence close to $1$ can be approached at multiple angles.
%This limit is approached at %$\gamma\mathrm{tan}(Z_{1,2})\to 0.5, A_2-A_1\to 0\phantom{1}\mathrm{or}\phantom{1}\pi$  $A_2-A_1\to\pi$ or $A_2-A_1\to 0,\gamma\mathrm{tan}(Z_{1,2})\to 0.6$. 
%We found there can be handful number of zenith angles $Z_{1,2}$ of emitted photons that can have satisfactory concurrence and double emission rate simultaneously.
%
However, for most regions in the double-angular parameter space, the concurrence is relatively low. 
%Thus, without selection or purification, the overall concurrence of the emitted photon pairs would have a small value of only 0.032 to 0.00. 
Thus, without selection or purification, the overall entanglement of the emitted photon pairs would vanish. 
%\tbaj{In the next section, we will show how to improve the concurrence as an entangled photon pair source.}

%\section*{Observation of entangled photon pairs}

\begin{figure}[thb]
    \centering
    \includegraphics[width = 80mm]{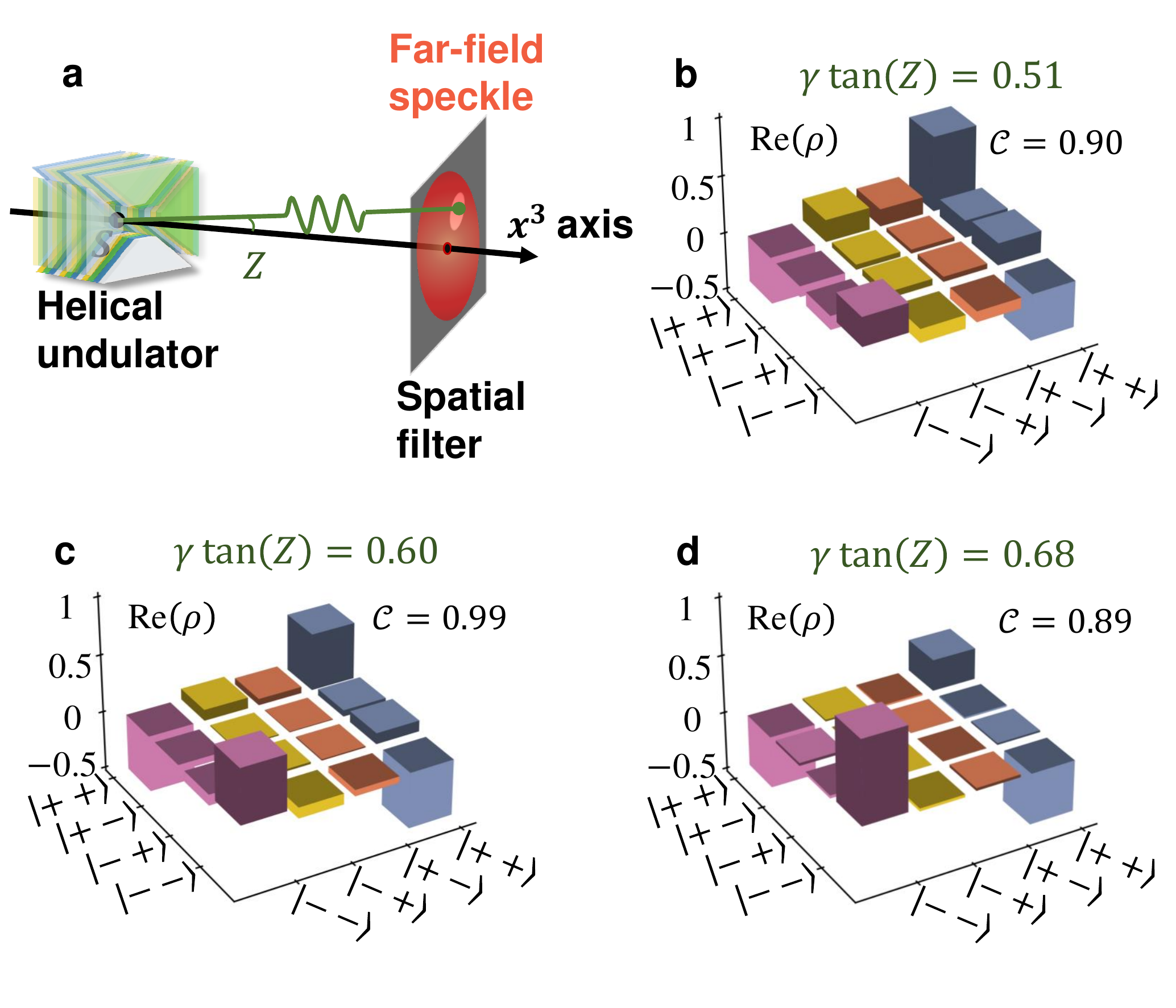}
    \caption{\newtxt{\textbf{Density matrices of the emitted photon pairs.} \textbf{a}, Geometry of the inserted circular aperture, where $Z$ describes the angular position of the circular hole relative to the equivalent source point $\mathrm{S}$. \textbf{b,c,d}, Real parts of the density matrices of the resulting entangled states for the aperture shown in \textbf{a} with $\gamma\mathrm{tan}(Z)=0.51,0.60,0.68$, which leads to $\mathcal{C}=0.90,0.99,0.89$, respectively. The size of the circular aperture is chosen to be $\sim 1/300$ of the far-field speckle. 
    %We also show the corresponding values of the effective coincidence ratio $r_{\mathrm{c}}$ and the coincidence count rate $R_{\mathrm{co}}$ for a $1~\mathrm{MHz}$ electron source. 
    The imaginary parts of the density matrices are zero due to the specific choice of the helicity basis and the identical emitting directions of the entangled photons (see Section VIII of SI for details).}}
    \label{fig:Conc_Puri}
\end{figure}

%For potential applications in X-ray quantum optics, we aim to produce entangled photon pairs at certain 2-qubit states. In Fig.~\ref{fig:Conc_Puri}a, we show the proposed experimental scheme for the observation of the entangled photon pair emission. A variable line spacing (VLS) grating spectrometer used in the spectral analysis of the FEL~\cite{brenner2011first,serkez2015soft} can diffract the incident FEL beam to different orders (see Section X of SI for details). 
%We can observe the 1st-order diffracted beams by placing two detectors with certain spectral response at certain positions along the focal plane.
%By placing two detectors with certain spectral response at certain positions along the focal plane, we can make a time-coincidence measurement of the entangled photon pairs in the 1st-order diffracted beams.

\newtxt{To obtain higher concurrence, a circular aperture can be inserted into the system at far field as shown in Fig.~\ref{fig:Conc_Puri}a, and only photon pairs that are emitted in the direction $Z_2\simeq Z_1\simeq Z$ can pass through the aperture. In light of the numerical calculation for the angular distribution of concurrence shown in Fig.~\ref{fig:Num_Res}b, the circular aperture can be chosen to maximize the concurrence of the photon pairs that pass it.
As demonstrated in Fig.~\ref{fig:Conc_Puri}b-d, concurrence close to $1$ can be achieved at multiple angles such as $\gamma\mathrm{tan}(Z)=0.51,0.60,0.68$ with a sufficiently small aperture, whose size is chosen to be $\sim 1/300$ of the far-field speckle.}

\newtxt{We can now estimate the emission rate of the entangled photon pairs in certain energy ranges that pass through the aperture in the lab frame. The energy ranges of the emitted photons are chosen to be $[\omega_{\mathrm{fd}}/3-10~\mathrm{eV},~\omega_{\mathrm{fd}}/3]$ and $[2\omega_{\mathrm{fd}}/3-10~\mathrm{eV},~2\omega_{\mathrm{fd}}/3]$. The angular position of the circular hole is chosen to be $\gamma\mathrm{tan}(Z)=0.60$. }

\newtxt{For a single electron, the emission rate of the entangled photon pairs is $R_{\mathrm{SE}}=6\times10^{-21}$. For an electron pulse in a synchrotron system, with the same amount of electrons $N_{\mathrm{e}}=1.12\times10^9$ as in Table~\ref{tab:ExpPara}, the total emission rate is $R_{\mathrm{Sync}}=N_{\mathrm{e}}R_{\mathrm{SE}}=7\times10^{-12}$. For a microbunched electron pulse in the FEL system where the number of electrons in each microbunch is chosen to be a typical value of $10^6$, with Eq.~\ref{eq:dWMb}, we obtain a rate of $R_{\mathrm{FEL}}=1.5\times10^{-4}$. The FEL system can thus provide an enhancement of order of $10^{7}$ for the entangled photon pair emission rate compared to the synchrotron system. 
%Under our scheme, an electron source with a repetition rate of $1~\mathrm{MHz}$ as in the design of LCLS-II-HE or EuXFEL~\cite{decking2020mhz} can produce entangled photon pairs with a rate of $150~\mathrm{s}^{-1}$. 
}

\section*{Conclusion}

In conclusion, we have theoretically demonstrated the entangled photon pair generation with FEL. In the electron frame, the double photon emission can be understood as the scattering between the electrons and the quasi-EM wave of the undulator field, which can be enhanced by the electron microbunching in FEL. \newtxt{We demonstrate that the FEL can generate
highly entangled photon pairs with enhanced emission rate, which is impossible from a classical perspective. The FEL can thus be utilized as an entangled photon pair source for X-ray quantum optics, especially for applications in the soft X-ray regime where it is challenging to use nonlinear effects like spontaneous parametric down conversion due to the strong absorption in nonlinear crystals.}

\section*{Methods}

We use the Floquet expansion of Volkov state in the the calculation of double emission scattering matrix $S_{\mathrm{fi}}$. As the dependence of the Volkov wave function on the phase $\phi$ (thus four-position $x$) shows a periodicity of $2\pi$, a Fourier expansion can be performed such that

\begin{equation}
    \Psi^{\mathrm{V}}_p(x)=\sum_{\tilde{n}=-\infty}^{+\infty}{F_{\tilde{n}} e^{-i(q\cdot x+\tilde{n} k\cdot x)}}u(p)\;.
\end{equation}
Thus, the contributions to $S_{\mathrm{fi}}$ can be divided according to the number of Floquet photons absorbed, $n$. E.g., for the Feynman diagram shown in Fig.~1c, 

\begin{equation}\label{eq:Sfi}
    \begin{split}
        S_{\mathrm{fi}} &=\int d^4x_1\int d^4x_2 \bar{\Psi}^{\mathrm{V}}_{\mathrm{f}}(x_2)(-ie\slashed{A}_2(x_2)) \\
        &\phantom{=}\times(iG_{\mathrm{V}}(x_2,x_1))(-ie\slashed{A}_1(x_1))\Psi^{\mathrm{V}}_{\mathrm{i}}(x_1) \\
        &\phantom{=}+\int d^4x_1\int d^4x_2 \bar{\Psi}^{\mathrm{V}}_{\mathrm{f}}(x_2)(-ie\slashed{A}_1(x_2)) \\
        &\phantom{=}\times(iG_{\mathrm{V}}(x_2,x_1))(-ie\slashed{A}_2(x_1))\Psi^{\mathrm{V}}_{\mathrm{i}}(x_1) \\
        &= -i\frac{e^2}{2V}\sqrt{\frac{1}{k_2^0k_1^0}}\sum_{n=-\infty}^{+\infty}{\bigg\{ \sum_{n_1=-\infty}^{+\infty}{M_{\mathrm{S}}(n,n_1)\bigg\vert_{\mathrm{MC}(n_1)}}} \\
        &\phantom{=}\times(2\pi)^4\delta^{(4)}(q_{\mathrm{f}}+k_1+k_2-q_{\mathrm{i}}-nk)\bigg\}\;,
    \end{split}
\end{equation}
where the helicity-relevant part $M_{\mathrm{S}}(n,n_1)\vert_{\mathrm{MC}(n_1)}$ is explained in Section II of SI, which only survives when momentum conservation is fulfilled at each vertex of the Feynman diagrams.  

The differential scattering rate is then calculated by considering the total scattering probability $\dot{W}$ of double emission  that satisfies

\begin{equation}\label{eq:dotW}
    \begin{aligned}
        \dot{W}=&\int{V\frac{d^3\boldsymbol{q}_{\mathrm{f}}}{(2\pi)^3}V\frac{d^3\boldsymbol{k}_1}{(2\pi)^3}V\frac{d^3\boldsymbol{k}_2}{(2\pi)^3}\frac{\vert S_{\mathrm{fi}}\vert^2}{VT}V} \\ =&\int\frac{4V^4}{(2\pi)^9}d^3\boldsymbol{q}_{\mathrm{f}}\int{d\Omega_{k_1}}\int{d\Omega_{k_2}}\int_{0}^{+\infty}{dk_1^0}\int_{0}^{+\infty}{dk_2^0}\,(k_1^0)^2(k_2^0)^2\frac{\vert S_{\mathrm{fi}}\vert^2}{VT}
        \;,
    \end{aligned}
\end{equation}
where $VT$ is the space-time volume of the interaction zone. % $VT=(2\pi)^4\delta^{(4)}(0)$ is considered infinite here.
%$dk_1^0,d\Omega_{k_1}$ and $d\Omega_{k_2}$, 
Combining Eq.~\ref{eq:Sfi} and Eq.~~\ref{eq:dotW}, and  completing the integral over variables except for $k_1^0$, $\Omega_{k_1}$ and $\Omega_{k_2}$, we obtain the Eq.~\ref{eq:ddotW} that

\begin{equation*}
    \begin{split}
        &\frac{d\dot{W}}{dk_1^0d\Omega_{k_1}d\Omega_{k_2}}=\frac{e^4}{(2\pi)^5}\frac{m^2}{2q_{\mathrm{i}}^0} \\
        &\phantom{=}\times\sum_{n=1}^{+\infty}{\frac{k_1^0(k_2^0)^2\Theta(q_{\mathrm{i}}^0+nk^0-k_1^0-k_2^0)M}{2\vert(q_{\mathrm{i}}+nk-k_1)\cdot k_2\vert}}\bigg\vert_{\mathrm{DE}(n)}\;,
    \end{split}
\end{equation*}
where $\vert_{\mathrm{DE}(n)}$ means the constraint of $ (q_{\mathrm{i}}+nk-k_1)\cdot k_2=q_{\mathrm{i}}\cdot nk-q_{\mathrm{i}}\cdot k_1-nk\cdot k_1$, and the helicity-relevant part $M$ is explained in Section II of SI. 

%%%%%%%%%%%%%%%
% reduction of the complicated entangled emission probability expression for better usage in experimental design.
% Following the referee's suggestion, we have obtained a reduced formula that clearly shows the dependence of the entangled emission probability on the key FEL parameters, see Eq.~(9) in the Methods Section of the revised main text.
\newtxt{In order to explicitly show the dependence of the entangled emission probability on the key FEL parameters, we can reduce the formula of the differential entangled double emission probability for a single electron in the lab frame based on Eq.~\ref{eq:ddotW}. Focusing on $n=1$, the entangled emission probability $U_{\mathrm{pair}}$ is given by

\begin{equation}
    U_{\mathrm{pair}}\simeq F_{\mathrm{L}}F_{\mathrm{S}}\frac{d\dot{W}}{dk_1^0d\Omega_{k_1}d\Omega_{k_2}}\bigg\vert_{\mathrm{EF},k_1,k_2}\;,
\end{equation}
where $\frac{d\dot{W}}{dk_1^0d\Omega_{k_1}d\Omega_{k_2}}\bigg\vert_{\mathrm{EF},k_1,k_2}$ is the value of scattering probability in Eq.~\ref{eq:ddotW} with the two photon four-momenta in EF chosen as $k_1,k_2$, and $F_{\mathrm{L}}$, $F_{\mathrm{S}}$ are the prefactors related to the Lorentz transformation and solid angle, respectively. 

%$U_{\mathrm{pair}}$ can be expressed with dimensionless quantities $\gamma$, $k^0/m$ and $K=ea/m$.
%, with coefficients of constant value at certain $k_1,k_2$. 
%Equivalently, 
$U_{\mathrm{pair}}$ can be expressed with key FEL parameters
$\gamma$, $\lambda_u$ and $B_0$, which represent the Lorentz factor of electrons, undulator period length and magnetic field amplitude of the undulator, in the form of

\begin{equation}\label{eq:Sc}
    \begin{aligned}
        U_{\mathrm{pair}}&\propto  \frac{Q(B_0\lambda_u)}{\lambda_u^2(1+(c_{\mathrm{K}}B_0\lambda_u)^2)^2(1+(c_{\mathrm{K}}B_0\lambda_u)^2/2)[1+(c_{\mathrm{K}}B_0\lambda_u)^2(1+\mathrm{cos}Z_2')/2]}\;,
    \end{aligned}
\end{equation}
where

\begin{equation}
    \begin{aligned}
        Q(B_0\lambda_u)&=\sum_{t_1,t_2,t_3,t_4=0}^{2}P_{t_1,t_2,t_3,t_4}(B_0\lambda_u) J_{t_1}(\mathrm{sin}Z_1'\frac{(c_{\mathrm{K}}B_0\lambda_u)}{1+(c_{\mathrm{K}}B_0\lambda_u)^2}) \\
        &\times J_{t_2}(\mathrm{sin}Z_1'\frac{(c_{\mathrm{K}}B_0\lambda_u)}{1+(c_{\mathrm{K}}B_0\lambda_u)^2})J_{t_3}(\mathrm{sin}Z_2'\frac{(c_{\mathrm{K}}B_0\lambda_u)}{1+(c_{\mathrm{K}}B_0\lambda_u)^2}) \\
        &\times J_{t_4}(\mathrm{sin}Z_2'\frac{(c_{\mathrm{K}}B_0\lambda_u)}{1+(c_{\mathrm{K}}B_0\lambda_u)^2})\;, \nonumber
        % &\simeq \frac{Q(B_0\lambda_u)}{\lambda_u^2(1+(c_{\mathrm{K}}B_0\lambda_u)^2)^2(1+(c_{\mathrm{K}}B_0\lambda_u)^2/2)[1+(c_{\mathrm{K}}B_0\lambda_u)^2(1+\mathrm{cos}Z_2')/2]}\;,
    \end{aligned}
\end{equation}
% and $U_{\mathrm{pair}}$ is reduced to
% \begin{eqnarray}
%         U_{\mathrm{pair}} \simeq \frac{Q(B_0\lambda_u)}{\lambda_u^2(1+(c_{\mathrm{K}}B_0\lambda_u)^2)^2(1+(c_{\mathrm{K}}B_0\lambda_u)^2/2)[1+(c_{\mathrm{K}}B_0\lambda_u)^2(1+\mathrm{cos}Z_2')/2]}\;,
% \end{eqnarray}
and $c_{\mathrm{K}}=e/2\pi m$. $P_{t_1,t_2,t_3,t_4}(B_0\lambda_u)$ are polynomials of $B_0\lambda_u$ with the highest order between 4 and 8, and $J_{t}(x)$ refers to the Bessel function of the first kind with order $t$. %In the non-perturbative region of $K\sim 1$, Eq.~\ref{eq:Sc} cannot be simplified anymore without numerical calculations. 
%And, $Q(B_0\lambda_u)$ is a integrated label that refers to the summation of polynomials multiplying Bessel functions. 
The detailed derivation of Eq.~\ref{eq:Sc} is presented in Section V of SI. 

%Numerical calculations on $Q(B_0\lambda_u)$ and thus $U_{\mathrm{pair}}$ is strongly recommended. Yet,
At the vicinity of a specific choice of $k_1, k_2$, we can obtain approximated power function for $Q(B_0\lambda_u)$. For example, consider $k_1,k_2$ that satisfy $\omega_1=\omega_{\mathrm{fd}}/3$ and $\gamma\mathrm{tan}Z_{1,2}\simeq0.51,0.60,0.68$ in the lab frame, with the corresponding $\omega_2\simeq2\omega_{\mathrm{fd}}/3$ given by the conservation of four-quasi-momentum. These three cases correspond to the three circumstances shown in Fig.~\ref{fig:Conc_Puri} of the main text. we obtain by fitting that

\begin{align}
    Q(B_0\lambda_u)\propto&(B_0\lambda_u)^{1.90}\quad\mathrm{for~~\gamma\mathrm{tan}Z_{1,2}\simeq0.51}\;, \\
    Q(B_0\lambda_u)\propto&(B_0\lambda_u)^{2.00}\quad\mathrm{for~~\gamma\mathrm{tan}Z_{1,2}\simeq0.60}\;, \\
    Q(B_0\lambda_u)\propto&(B_0\lambda_u)^{2.10}\quad\mathrm{for~~\gamma\mathrm{tan}Z_{1,2}\simeq0.68}\;.
\end{align}
The details are shown in Section V of SI.
}
%%%%%%%%%%%%%%%

%We choose the undulator configuration to be helical rather than linearly polarized according to the following reason. %The rotational symmetry about the $x^3$ axis 
The helical undulator simplifies the dependence of the double-emission process on the emitting azimuth angles $A_1,A_2$ to $A_2-A_1$ and reduces the parameter space, making it more convenient to optimize the detection scheme for the entangled photon pairs in experiment. A detailed calculation for the case of linearly polarized undulator is given in Section VI in SI.

\section*{Data availability}
The data that support the findings of this study are available from the corresponding authors upon reasonable request.

\section*{Code availability}
The code for the simulation is available from the corresponding authors upon reasonable request.

\section*{Acknowledgement}
The authors are thankful to Zhirong Huang and Zhaoheng Guo for helpful discussions. This work was supported by the National Natural Science Foundation of China (No. 12174009, 11974031, 12174011).

\end{document}